\documentclass[reprint,superscriptaddress,prb,longbibliography]{revtex4-2}

\usepackage{color}
\usepackage{graphicx}
\usepackage{dcolumn}
\usepackage{bm}
\usepackage{hyperref}
\hypersetup{colorlinks=true,linkcolor=blue,citecolor=blue,urlcolor=blue}

\usepackage[T1]{fontenc} 
\usepackage{newtxtext}   
\usepackage{newtxmath}   
\usepackage{upgreek}
\usepackage{mhchem}

\begin{document}

\title{Impact of Thermal Effects on the Current-Tunable Electrical Transport in the Ferrimagnetic Semiconductor Mn$_3$Si$_2$Te$_6$}  

\author{Yiyue Zhang}
\thanks{These authors contributed equally to this work. }
\affiliation{Low Temperature Physics Laboratory, College of Physics \& Center of Quantum Materials and Devices, Chongqing University, Chongqing 401331, China}

\author{Xin Jin}
\thanks{These authors contributed equally to this work. }
\affiliation{College of Physics and Electronic Engineering, Chongqing Normal University, Chongqing 401331, China}   

\author{ZeYu Li}
\thanks{These authors contributed equally to this work. }
\affiliation{Low Temperature Physics Laboratory, College of Physics \& Center of Quantum Materials and Devices, Chongqing University, Chongqing 401331, China}

\author{Kunya Yang}
\thanks{These authors contributed equally to this work. }
\affiliation{Low Temperature Physics Laboratory, College of Physics \& Center of Quantum Materials and Devices, Chongqing University, Chongqing 401331, China}

\author{Linlin Wei}
\thanks{These authors contributed equally to this work. }
\affiliation{Low Temperature Physics Laboratory, College of Physics \& Center of Quantum Materials and Devices, Chongqing University, Chongqing 401331, China}

\author{Xinrun Mi}

\affiliation{Low Temperature Physics Laboratory, College of Physics \& Center of Quantum Materials and Devices, Chongqing University, Chongqing 401331, China}

\author{Aifeng Wang}
\affiliation{Low Temperature Physics Laboratory, College of Physics \& Center of Quantum Materials and Devices, Chongqing University, Chongqing 401331, China}

\author{Xiaoyuan Zhou}
\affiliation{Low Temperature Physics Laboratory, College of Physics \& Center of Quantum Materials and Devices, Chongqing University, Chongqing 401331, China}

\author{Xiaolong Yang}
\email{yangxl@cqu.edu.cn}
\affiliation{Low Temperature Physics Laboratory, College of Physics \& Center of Quantum Materials and Devices, Chongqing University, Chongqing 401331, China}

\author{Yisheng Chai}
\email{yschai@cqu.edu.cn}
\affiliation{Low Temperature Physics Laboratory, College of Physics \& Center of Quantum Materials and Devices, Chongqing University, Chongqing 401331, China}

\author{Mingquan He}
\email{mingquan.he@cqu.edu.cn}
\affiliation{Low Temperature Physics Laboratory, College of Physics \& Center of Quantum Materials and Devices, Chongqing University, Chongqing 401331, China}

\date{\today}

\begin{abstract}

In the ferrimagnetic semiconductor Mn$_3$Si$_2$Te$_6$, a colossal magnetoresistance (CMR) is observed only when a magnetic field is applied along the magnetic hard axis ($\mathbf{H}\parallel c$). This phenomenon suggests an unconventional CMR mechanism potentially driven by the interplay between magnetism, topological band structure, and/or chiral orbital currents (COC). By comparing electrical resistance measurements using continuous direct currents and pulse currents, we found that the current-induced insulator-metal transition, supporting the COC-driven CMR mechanism, is likely a consequence of Joule heating effects. First-principles calculations reveal a pronounced band gap reduction upon tilting the magnetic moments toward the $c$-axis, accompanied by increased carrier concentration and Fermi velocity. Combining spin orientation-dependent electronic structure with Boltzmann transport theory, the calculated electrical resistance closely reproduces the CMR observed experimentally. These findings suggest that the CMR in Mn$_3$Si$_2$Te$_6$ stems primarily from band gap reduction induced by partial polarization of magnetic moments along the magnetic hard axis. 

\end{abstract}

\maketitle

The interplay between charge, spin, and lattice degrees of freedom in magnetic systems can give rise to phenomena like giant and colossal magnetoresistance (GMR and CMR). CMR, observed in materials such as \ce{La_{1–x}Ca_{x}MnO3} \cite{APRamirez_1997,Tokura_2006} and \ce{Tl2Mn2O7} \cite{Shimakawa1996TlMnO,SubramanianTlMnO,RamirezTlMnO}, is characterized by large resistance changes—up to several orders of magnitude—under a magnetic field. This property makes CMR materials promising for high-density storage applications. In doped manganites like \ce{La_{1–x}Ca_{x}MnO3}, CMR arises from complex mechanisms such as double-exchange interactions, Jahn-Teller distortions, and charge/orbital ordering \cite{ZenerDEI,GennesDEI,AndersonDEI}. In \ce{Tl2Mn2O7}, CMR is thought to result from interactions between conduction electrons and spin fluctuations or magnetic polarons \cite{Shimakawa1996TlMnO,SubramanianTlMnO,RamirezTlMnO}. A common feature of CMR materials is that aligning magnetic moments under a field reduces electron scattering or band gaps, leading to lower electrical resistance.

\begin{figure}
\includegraphics[width=220pt]{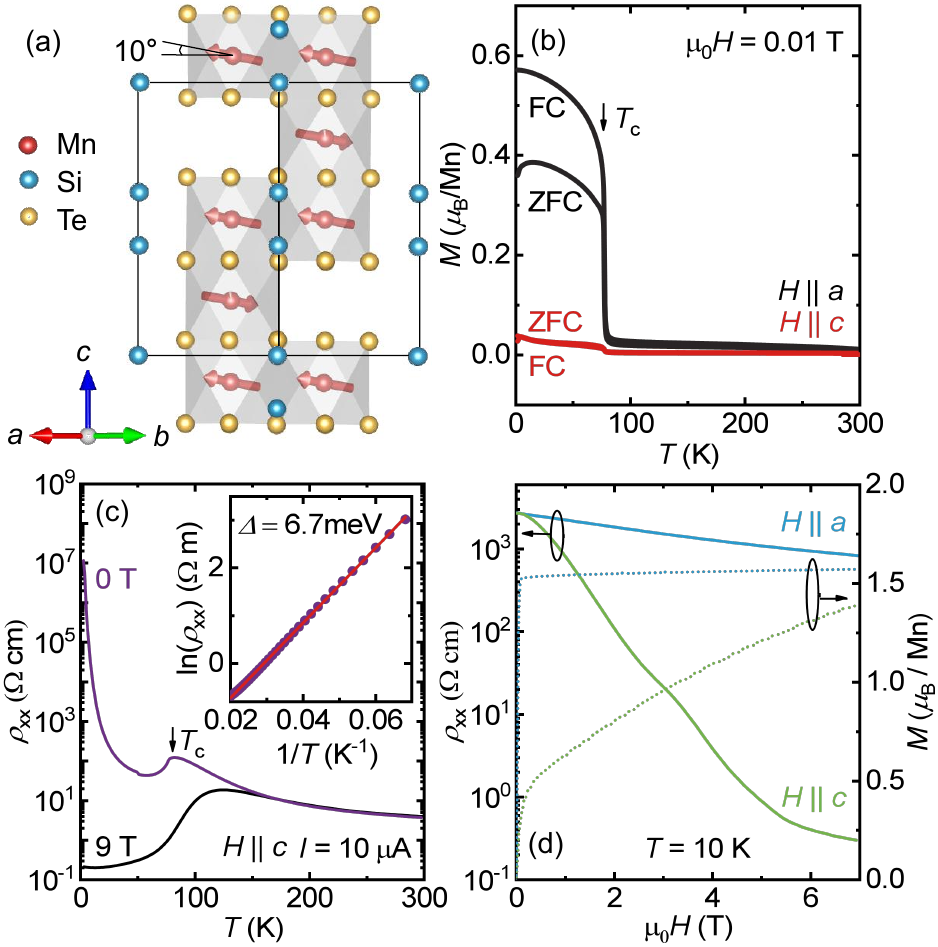}
\caption{\label{fig:1}(a) Side view of the crystal structure of \ce{Mn3Si2Te6}, with Mn moments (red arrows) tilted about 10° from the $ab$-plane below $T_\mathrm{c} = 78$ K. (b) Temperature-dependent magnetization measured in 0.01 T using zero-field-cooled (ZFC) and field-cooled (FC) protocols for magnetic fields along the $a$- and $c$-axes.   (c) Temperature dependence of resistivity ($\rho_{xx}$) in 0 and 9 T ($H\parallel c$) with a small current ($I = 10$ µA) applied along the $a$-axis. Inset: Linear scaling of ln($\rho_{xx}$) versus $1/T$ between 15 K and 50 K. (d) Comparison of magnetization and resistivity at 10 K for magnetic fields applied along different directions.}
\end{figure}

Recent studies have revealed unique colossal magnetoresistance (CMR) characteristics in the ferrimagnetic semiconductor \ce{Mn3Si2Te6}, differing from conventional behavior.  As shown in Fig.~\ref{fig:1}(a), \ce{Mn3Si2Te6} adopts a trigonal $P$$\bar{3}$1$c$ structure, where edge-sharing \ce{MnTe6} octahedra form alternating honeycomb (Mn1) and triangular (Mn2) layers along the $c$-axis \cite{VINCENT1986349,RIMET19817,may2017magnetic,may2020magnetic}. Below the ferrimagnetic transition temperature ($T_\mathrm{c} \sim$ 78 K), magnetic moments on both Mn1 and Mn2 sites order ferromagnetically, primarily within the $ab$-plane. Antiferromagnetic coupling between the sublattices results in a ferrimagnetic state. Notably, a drastic negative CMR, up to nine orders of magnitude, occurs only when the magnetic field is applied along the hard $c$-axis, contrasting sharply with the moderate negative MR seen when the field is applied in the easy $ab$-plane. This highlights a significant departure from typical CMR materials \cite{ni2021colossal, zhang2022control, seo2021colossal, wang2022pressure}.

\begin{figure*}
\includegraphics[width=350pt]{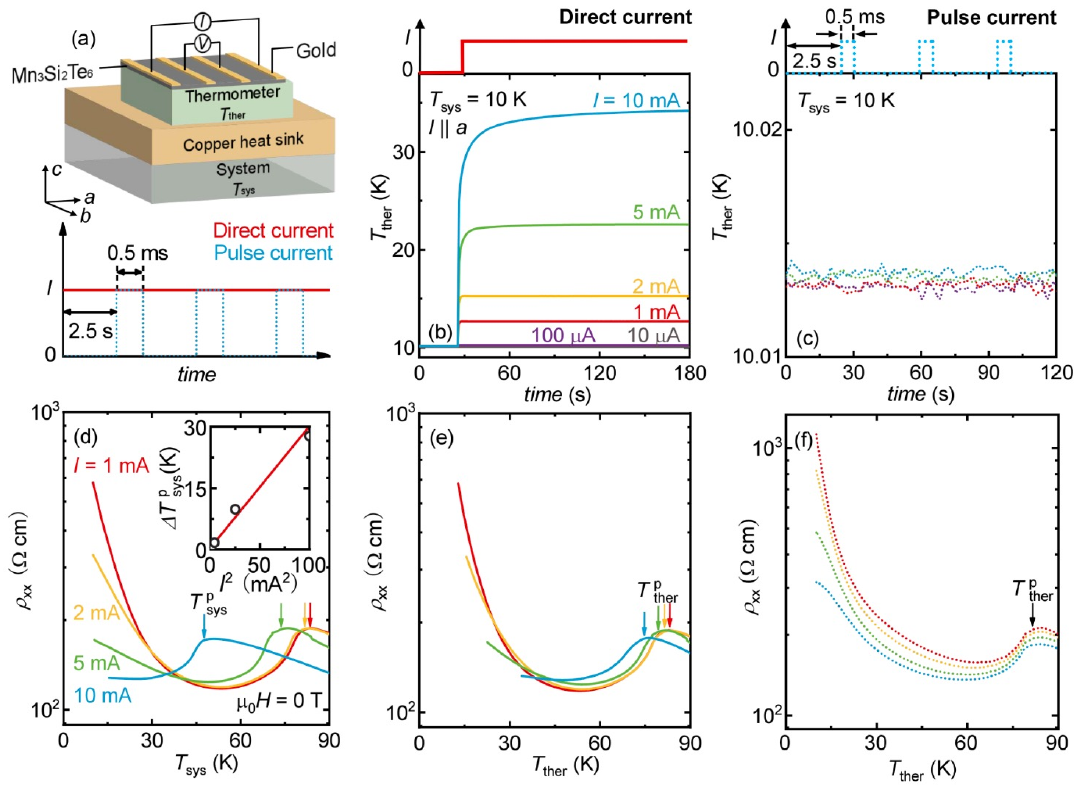}
\caption{\label{fig:2}Upper panel: Experimental configuration for electrical transport measurements, showing the \ce{Mn3Si2Te6} sample thermally anchored to both a thermometer and copper heat sink. Lower panel: Comparison of current profiles for pulsed (blue dashed) and direct (red solid) measurements. (b,c) Current-dependent temperature evolution recorded by $T_\mathrm{ther}$ at fixed $T_\mathrm{sys}=10$ K. (d,e) Resistivity ($\rho_{xx}$) versus temperature in DC mode, plotted against (d) $T_\mathrm{sys}$ and (e) $T_\mathrm{ther}$.  Inset in (d) shows the linear relationship between $\Delta T^\mathrm{p}_\mathrm{sys}$ and $I^2$. (f) Temperature-dependent $\rho_{xx}$ measured in pulse current mode, plotted versus  $T_\mathrm{ther}$.}  
\end{figure*}

Two main scenarios have been proposed to explain the unconventional CMR observed in \ce{Mn3Si2Te6}. The first suggests that CMR arises from the reduction of the band gap due to the lifting of spin orientation-dependent nodal-line band degeneracy \cite{seo2021colossal, Zhang_nodal, Susilo2024}. In this scenario, when the spins lie within the $ab$-plane ($M \parallel a$), a finite band gap exists, and the spin-polarized Te valence bands exhibit a twofold nodal-line degeneracy near the Fermi energy. The band gap closes when the spins align along the $c$-axis ($M \parallel c$), driving an insulator-to-metal transition and negative CMR \cite{seo2021colossal, Zhang_nodal, Susilo2024}.  However, the CMR appears most pronounced at magnetic fields where spins tilt only from about 10$^\circ$ to about 30$^\circ$ toward the  $c$-axis \cite{Feng_nonlinear},  raising questions about whether such partial alignment suffices for complete gap closure. 

An alternative explanation involves electron scattering due to COC-induced orbital magnetic moments ($M_\mathrm{COC}$) \cite{zhang2022control, Zhang2023_hall,Zhang_current,Lovesey_COC}. In this model, $M_\mathrm{COC}$, which orients along the $c$-axis, is induced by chiral orbital currents running along the Te-Te edges in the $ab$-plane of \ce{MnTe6} octahedra. When a magnetic field is applied along the $c$-axis, it aligns these randomly distributed $M_\mathrm{COC}$ domains, reducing electron scattering and lowering resistance. The COC-driven CMR mechanism is supported by electric current tunable properties including current-induced suppression of resistivity and $T_\mathrm{c}$, and a first-order-like transition in the current-voltage ($I$-$V$) characteristics \cite{zhang2022control, Zhang2023_hall, Zhang_current}. However, Joule heating effects could mimic these phenomena, particularly given the samples' high resistance ($\sim 10^3$ $\Omega$ at 10 K for typical $1 \times 0.5 \times 0.1$ mm$^3$ sized crystals) and the mA-scale currents required for the current control of the COC state.

In this work, we first investigate the influence of thermal effects on the current-tunable electrical transport properties by directly attaching a thermometer to the sample and comparing electrical transport behavior under continuous DC and pulsed currents. Our results indicate that Joule heating plays a significant role in the observed current-dependent electrical transport properties. Second, we perform direct calculations of the electrical resistance using the spin orientation-dependent electronic band structure. The calculated resistance well reproduces the experimentally observed CMR, demonstrating that field-induced band gap reduction is the dominant mechanism for CMR in \ce{Mn3Si2Te6}.   Detailed experimental methods are provided in the Supplemental Materials \cite{supplemental} (see also references \cite{Ran2023,Chai2021,zhang2023,PRB1994Blohl,PRB1999Kresse,PRB1993Kresse,PRB1994Kresse,PRB1996Kresse,PRL1996Perdew,PRB1991U,JPCM2003U,PRB2000U,Zhang_nodal,MADSEN2018140,Feng_nonlinear,may2017magnetic} therein).

\begin{figure}
\includegraphics[width=225pt]{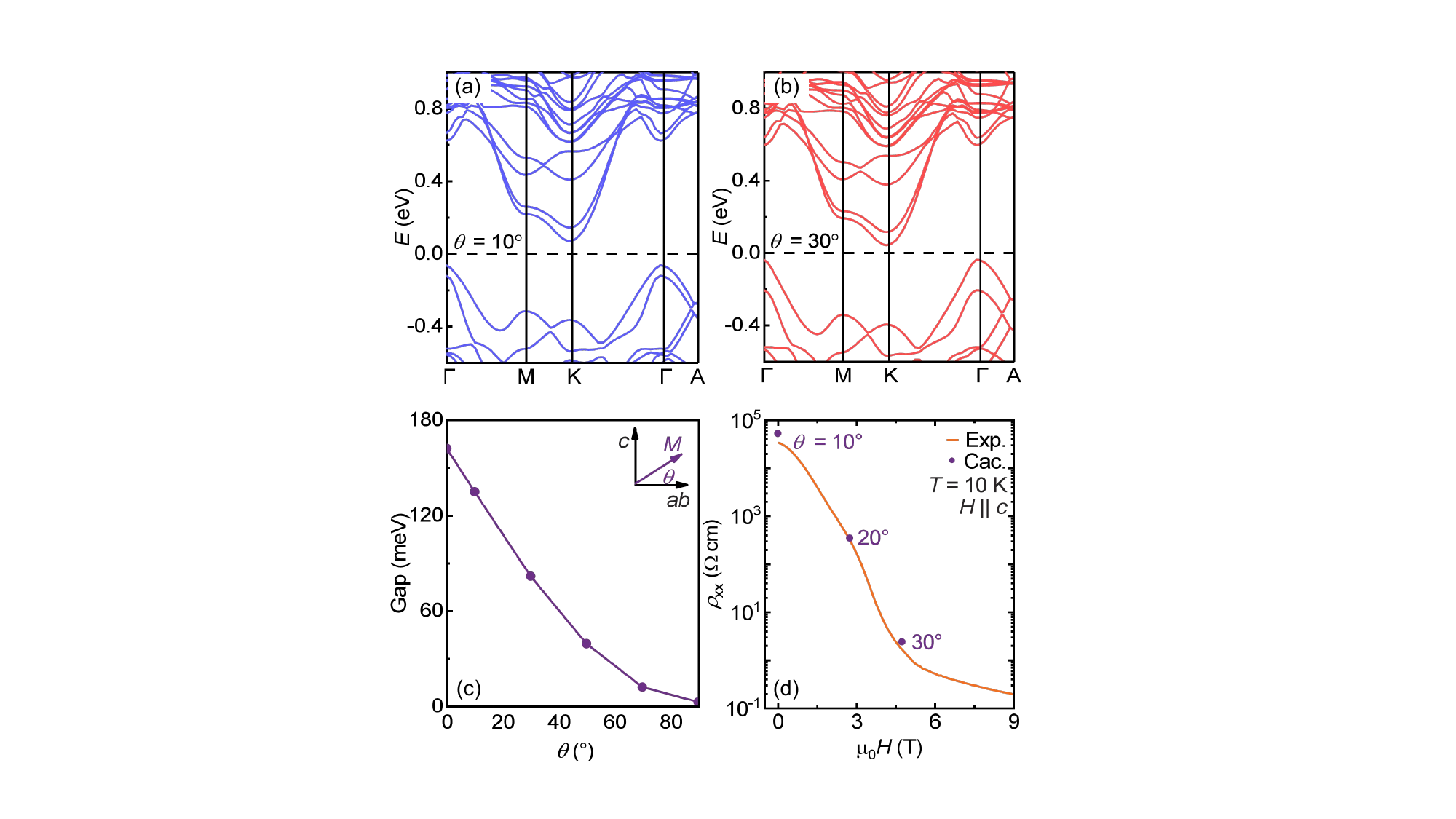}
\caption{\label{fig:4} Electronic band structure calculations for spin orientations tilted (a) $10^\circ$ and (b) $30^\circ$  toward the $c$-axis. (c) Calculated band gap versus tilt angle $\theta$ (illustrated schematically in inset). (d) Comparison between experimental resistivity (orange solid curve) and theoretical calculations (purple data points) based on spin orientation-dependent band structure. }
\end{figure}

Figure ~\ref{fig:1}(b) shows the temperature-dependent magnetization of a \ce{Mn3Si2Te6} single crystal, with a ferrimagnetic transition at $T_\mathrm{c} = 78$ K, consistent with previous reports \cite{may2017magnetic, may2020magnetic, ni2021colossal, seo2021colossal, Ran2023, liu2021polaronic,Liuy2018,Bigi2023}. In the ferrimagnetic state, the in-plane magnetization ($H \parallel a$) is much larger than the out-of-plane magnetization ($H \parallel c$), indicating an in-plane easy axis \cite{may2017magnetic}. Figure ~\ref{fig:1}(c) shows the temperature-dependent resistivity ($\rho_{xx}$) of a \ce{Mn3Si2Te6} sample with a current applied along the $a$-axis ($I \parallel a$). In zero magnetic field, $\rho_{xx}$ shows semiconducting behavior, with a sharp drop below $T_\mathrm{c}$, likely due to the suppression of fluctuating short-range magnetic order as the material transitions to long-range ferrimagnetism \cite{liu2021polaronic, ni2021colossal, seo2021colossal, Ran2023}. Fitting the resistivity data between 15 and 50 K with a thermally activated model $\rho_{xx}(T) = \rho_0 \exp(\Delta / k_\mathrm{B} T)$ yields an energy gap of $\Delta = 6.7(3)$ meV, in agreement with earlier studies \cite{seo2021colossal, Zhang_nodal, may2017magnetic, Ran2023}. This gap is smaller than the $\sim$130 meV gap calculated for Mn moments tilted 10$^\circ$ from the $ab$-plane [see Fig. \ref{fig:4}], suggesting the presence of impurity bands in real \ce{Mn3Si2Te6} samples \cite{seo2021colossal}. When a magnetic field is applied along the $c$-axis, the resistivity decreases significantly, with CMR reaching $\sim10^8$ at 2 K in 9 T. The negative magnetoresistance extends up to about 150 K, likely due to field-induced suppression of fluctuating short-range magnetic order in the paramagnetic phase \cite{ni2021colossal}. These results indicate that the MR behavior in \ce{Mn3Si2Te6} is closely linked to spin degrees of freedom.

Figure ~\ref{fig:1}(d) compares the magnetization and magnetoresistance (MR) measured at 10 K with magnetic fields applied along different directions. For $H \parallel a$, the magnetization $M_\mathrm{a}$ saturates at $\sim$ 1.6 $\mu_\mathrm{B}$/Mn in a small field of $\sim$ 0.1 T. In contrast, for $H \parallel c$, saturation is not reached even in 7 T, and magnetic fields above 13 T are required for $M_\mathrm{c}$ to reach 1.6 $\mu_\mathrm{B}$/Mn \cite{ni2021colossal, seo2021colossal}. The sharp rise in $M_\mathrm{c}$ below 1 T results from the alignment of magnetic domains \cite{may2017magnetic, Feng_nonlinear}. Notably, CMR is observed only when the magnetic field is applied along the magnetic hard $c$-axis. In contrast, only moderate negative MR is observed when the field is applied within the easy $ab$-plane. Furthermore, within the field range ($\sim$ 5 T) where CMR occurs, the magnetic moment is far from saturation, distinguishing the CMR behavior in \ce{Mn3Si2Te6} from other CMR materials where magnetic polarization is essential.

Figure ~\ref{fig:2} explores the possible origin of current control over the electrical properties of \ce{Mn3Si2Te6}. In typical commercial cryostats, such as the Physical Property Measurement System (PPMS), the system thermometer is often placed away from the sample, leading to inaccurate determination of the sample temperature, especially at low temperatures and with large electric currents. To better estimate the sample temperature ($T_\mathrm{samp}$), we directly attached a Lakeshore Cernox-SD thermometer ($3\times2\times1$ mm$^3$) to the \ce{Mn3Si2Te6} sample, which was thermally anchored to a copper heat sink [Fig. \ref{fig:2}(a)].The Cernox thermometer was carefully calibrated prior to measurements, with detailed procedures provided in Supplemental Material the \cite{supplemental}. However, due to the low thermal conductivity of \ce{Mn3Si2Te6} \cite{liu2021polaronic}, a finite temperature gradient can still cause $T_\mathrm{samp}$ to be significantly higher than $T_\mathrm{ther}$ when Joule heating is significant. To minimize contact resistance, a thin layer of gold was sputtered onto the sample before attaching silver epoxy to the current and voltage leads. At 10 K, the resistance of the current leads is about 50 k$\Omega$. The dominant heat capacity of the setup comes from the copper heat sink (at 10 K, $C_\mathrm{Cu}\approx$ 1 mJ/K \cite{Cu_capa}) due to its substantially larger mass (1 g) compared to both the sample (3 mg) and thermometer (28 mg). A simple estimate suggests that a 1 mA current applied for 1 s would raise the temperature by $\sim$50 K, if heat dissipation and heat conduction to the surroundings are ignored. Fig.~\ref{fig:2}(b) shows that at $T_\mathrm{sys}=10$ K, a 1 mA DC current elevates $T_\mathrm{ther}$ to 12.5 K at thermal equilibrium.  Larger currents produce more severe heating, with $T_\mathrm{ther}$ reaching 34.2 K at 10 mA. Even at $T_\mathrm{sys}=90$ K, a 10 mA DC current causes a 12 K increase in $T_\mathrm{ther}$ (see Supplemental Material Fig. S1 \cite{supplemental}). In contrast, pulse current measurements (0.5 ms pulses at 0.4 Hz repetition rate) show negligible heating effects [Fig.~\ref{fig:2}(c)]. The pulse duration is three orders of magnitude shorter than the system's thermal relaxation time ($\sim$30 s at 10 K), effectively eliminating thermal artifacts.

Figure~\ref{fig:2}(d-f) compares the effects of DC and pulse currents on electrical resistivity, measured during warming at 1 K/min. Isothermal and temperature-ramped measurements show no significant differences (see Supplemental Material Fig. S2 \cite{supplemental}). However, notable discrepancies emerge between forward ($I>0$) and backward ($I<0$) DC current measurements due to thermoelectric effects, particularly at high currents (Fig. S2 \cite{supplemental}). To minimize these effects, we averaged resistances obtained from both current polarizations.  When plotted against system temperature ($T_\mathrm{sys}$), the electrical resistivity is significantly suppressed with increasing current. Below 30 K, a current-induced insulator-metal transition is seemly observed, along with a downward shift of the 'magnetic transition temperature' near 80 K - phenomena previously attributed to chiral orbital current (COC) states \cite{zhang2022control, Zhang2023_hall, Zhang_current}. However, substantial Joule heating at high DC currents creates a large discrepancy between $T_\mathrm{sys}$ and $T_\mathrm{samp}$. In this case, the magnetic transition temperature can not be accurately determined from $T_\mathrm{sys}$. We therefore define $T^\mathrm{p}_\mathrm{sys}$ as the resistivity maximum near 80 K to track the current-induced temperature shift $\Delta T^\mathrm{p}_\mathrm{sys} = T^\mathrm{p}_\mathrm{sys}(0) - T^\mathrm{p}_\mathrm{sys}(I)$. Not surprisingly, $\Delta T^\mathrm{p}_\mathrm{sys}$ follows an $I^2$ dependence [Fig.~\ref{fig:2}(d) inset], indicating heating-dominated behavior.  In contrast, plotting resistivity against $T_\mathrm{ther}$ [Fig.~\ref{fig:2}(e)] reveals much weaker current effects, though residual $\Delta T^\mathrm{p}_\mathrm{ther}\sim5$ K at 10 mA persists due to finite gradients between $T_\mathrm{ther}$ and $T_\mathrm{samp}$. Most strikingly, pulse current measurements show negligible temperature shifts even at 10 mA, with no insulator-metal transition and minimal current-induced suppression in resistivity [Fig.~\ref{fig:2}(f)]. These results demonstrate that Joule heating dominates the apparent current-tunable properties in DC measurements where accurate sample temperature determination is challenging. Note that, compared to the DC current mode, Joule heating effects are weaker under pulsed current conditions. As a result, the low-temperature resistivity measured in the pulsed current mode is higher than that measured in the DC mode, even at a modest current of 1 mA. Nevertheless, finite Joule heating effects are still present in the pulsed case, and a noticeable suppression of resistivity can be observed at low temperatures when large pulsed currents are applied. This behavior arises from nonzero temperature gradients between the thermometer temperature ($T_\mathrm{ther}$) and the actual sample temperature ($T_\mathrm{samp}$).

\begin{figure}
\includegraphics[width=250pt]{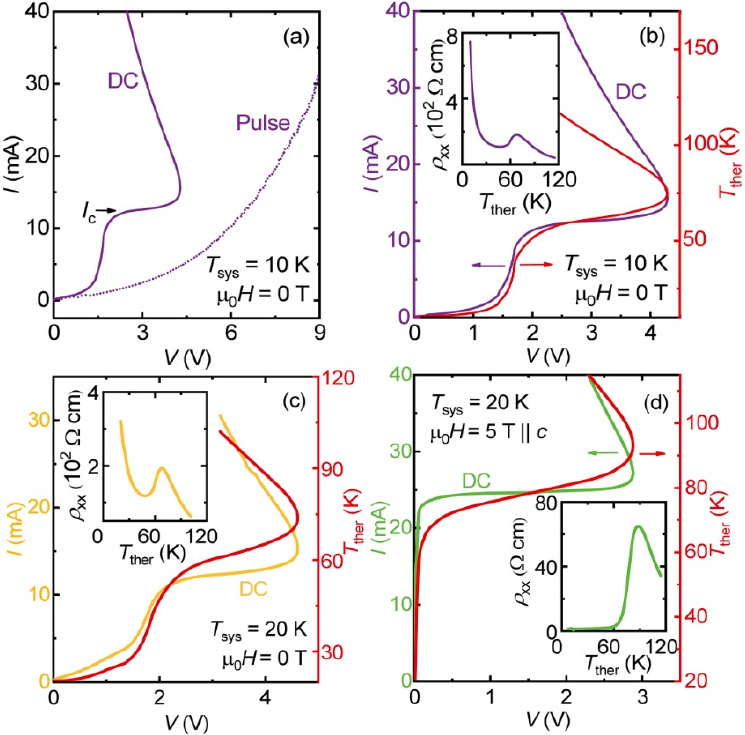}
\caption{\label{fig:3} (a) Current-voltage characteristics ($I-V$) at $T_\mathrm{sys} = 10$ K comparing direct (solid purple) and pulsed (dashed purple) current modes. (b) and (c) Comparison of $I-V$ curve with voltage-dependent $T_\mathrm{ther}$ at $T_\mathrm{sys} = 10$ and 20 K in the direct current mode and in zero magnetic field. (d) $I-V$ and $T_\mathrm{ther}-V$ curves measured at $T_\mathrm{sys} = 20$ K and in 5 T ($H\parallel c$). Insets in (b-d): Derived resistivity obtained from $I-V$ data plotted versus $T_\mathrm{ther}$, demonstrating the thermal origin of transport anomalies.}
\end{figure}

Distinct $I-V$ characteristics are observed in DC and pulse current modes, as presented in Fig. \ref{fig:3}. In DC mode, an unusual first-order-like transition is seen at a 'critical current' $I_\mathrm{c} \sim 10$ mA, similar to previous studies \cite{zhang2022control, Zhang2023_hall}, while a typical semiconductor-like $I-V$ behavior is observed in pulse mode. In DC mode, $T_\mathrm{ther}$ rises significantly due to Joule heating [Fig. \ref{fig:3}(b)]. Notably, the anomalous $I-V$ curve follows the $T_\mathrm{ther} - V$ curve. When the resistance obtained from the $I-V$ curve ($R = V/I$) is plotted against $T_\mathrm{ther}$, the temperature-dependent resistance of \ce{Mn3Si2Te6} is recovered. This behavior persists at $T_\mathrm{sys}=20$ K and under applied $c$-axis magnetic fields [Fig. \ref{fig:3}(c,d)]. Rapid rise in voltage near $I_\mathrm{c}$ corresponds to Joule heating driving the sample through the resistance maximum near $T_\mathrm{p}$, where resistance increases rapidly. Magnetic fields suppress resistance through the CMR effect, consequently requiring higher $I_\mathrm{c}$ to achieve the same thermal transition [Fig. \ref{fig:3}(d)]. Pulse measurements show minimal heating effects and exhibit normal transport behavior throughout the measured current range. These observations demonstrate that the apparent first-order transition in DC mode is strongly influenced by thermal effects.

The proposed chiral orbital current (COC) mechanism for colossal magnetoresistance (CMR) in \ce{Mn3Si2Te6} has been indirectly supported by current-dependent electrical and magnetic properties \cite{zhang2022control,Zhang2023_hall}. However, as demonstrated above, Joule heating effects dominate under large applied currents, especially at low temperatures, in this semiconducting system. Direct experimental evidence for the COC state warrants further study under rigorously controlled thermal conditions. Instead, our findings strongly support an alternative mechanism: field-induced band gap suppression \cite{seo2021colossal,Susilo2024,Zhang_nodal}.  This scenario is consistent with observations in other materials like \ce{EuMnSb2} \cite{Sun2021} and  \ce{EuTe2} \cite{YinEuTe2}, where spin orientation-dependent band gaps explain CMR. For \ce{Mn3Si2Te6}, neutron diffraction shows that magnetic moments tilt only modestly toward the $c$-axis (from $\sim10^{\circ}$ to $\sim30^{\circ}$ within $\sim$5 T)  \cite{Feng_nonlinear}. Nevertheless, our calculations reveal a substantial band gap reduction of ~50 meV [see Fig. \ref{fig:4}(a-c)], in agreement with prior theoretical work \cite{Zhang_nodal}. This reduction exceeds the measured gap (~7 meV) in actual samples.

To quantitatively connect band structure changes to transport properties, we calculated the electrical conductivity ($\sigma_{xx}$) by solving the Boltzmann transport equation based on the spin orientation-dependent band structure \cite{CPC2006,CPC2018}:
\begin{equation}
\sigma_{xx} = e^2 \int \sum_{\lambda} v^2_{\lambda} \tau_{\lambda} \left( -\frac{\partial f_0}{\partial \varepsilon} \right) d\varepsilon,
\end{equation}
where $e$, $v_\lambda$, $\tau_\lambda$, $f_0$, and $\varepsilon$ denote electronic charge, group velocity, relaxation time, Fermi-Dirac distribution, and energy eigenvalue of the carrier, respectively. For the calculations at fixed temperatures, we employed a constant relaxation time approximation with $\tau_\lambda=$100 fs. The chemical potential ($\mu-E_\mathrm{F}=0.136$ eV at 10 K) was fixed to ensure the calculated zero-field resistivity matched experimental values. This approach yielded a carrier concentration of $\sim10^{18}$ cm$^{-3}$, consistent with our previous experimental measurements on similar samples \cite{Ran2023} and prior work reported in \cite{ni2021colossal}.  Tilting spins from 10$^\circ$ to 30$^\circ$  increases both Fermi velocity  ($1.258 \times 10^{5}$ to $1.700 \times 10^{5}$~m/s) and the carrier concentration ($3.105 \times 10^{18}$ to $9.880 \times 10^{18}$~cm$^{-3}$), thereby significantly enhancing $\sigma_{xx}$. Note that, in general, a higher Fermi velocity corresponds to a lower density of states (DOS) at the Fermi level, which would typically lead to a reduced carrier density if the band edge remains fixed. In our case, however, as the spin tilting angle increases, the conduction band minimum shifts downward while the chemical potential remains fixed at a given temperature. This band-edge movement results in a net increase in carrier density. The resulting resistivity ($\rho_{xx} = 1/\sigma_{xx}$) closely reproduces the experimentally observed CMR [see Fig.~\ref{fig:4}(d)], highlighting the dominant role of the spin orientation-dependent band structure in governing the transport behavior. Although contributions from impurity levels are not captured in our calculations based on the intrinsic electronic structure, they are unlikely to be the primary driving force.  The Hall conductivity ($\sigma_{xy}$) is few orders of magnitude smaller than $\sigma_{xx}$ \cite{Ran2023,ni2021colossal}, and is ignored in the calculation of $\rho_{xx}$. The magnetic field-dependent spin tilting angle is inferred from previous neutron diffraction experiments \cite{Feng_nonlinear}.  In addition, electron scattering by short-range magnetic order may contribute to the moderate MR above $T_\mathrm{c}$ \cite{Feng_nonlinear}. However, these minor effects can unlikely cause the CMR in the magnetically long-range ordered state. In addition, field-induced magnetic transitions can also cause CMR, as found in manganite Sm$_{0.5}$Ca$_{0.25}$Sr$_{0.25}$MnO$_3$ \cite{Banik2018}.  Although few field-induced metamagnetic transitions are resolved in \ce{Mn3Si2Te6} by ac magnetostriction experiments, the CMR occurs primarily within the ferrimagnetic ground state [see Figs. S6 and S7 in Supplemental Material \cite{supplemental}], suggesting minor contributions from field-induced transitions.     

In summary, we examined Joule heating effects in the current-controlled electrical transport properties of \ce{Mn3Si2Te6}. Comparative studies using pulsed and direct current measurements reveal that Joule heating dominates the observed phenomena in DC mode. The band gap decreases substantially even though the magnetic moments only tilt slightly towards the $c$-axis. This is accompanied by an increase in both carrier concentration and Fermi velocity, which collectively contribute to the reduction of electrical resistance. Crucially, our first-principles calculations, combined with Boltzmann transport theory, closely reproduce the experimental CMR based on spin-dependent band structure. This agreement underscores the pivotal role of field-tunable electronic structure in the CMR of \ce{Mn3Si2Te6}.   

\medskip

\textit{Note added}: Recently, we became aware of recent experimental \cite{Fangthermal}  and theoretical studies \cite{liu2025frame} that corroborate the dominant role of Joule heating in current-controlled magnetic and transport properties of \ce{Mn3Si2Te6}.

\medskip

We thank Quansheng Wu for fruitful discussions. We thank Guiwen Wang and Yan Liu at the Analytical and Testing Center of Chongqing University for technical support. This work has been supported by National Natural Science Foundation of China (Grant No. 12474141), the Open Fund of the China Spallation Neutron Source Songshan Lake Science City (Grant No. KFKT2023B07), Fundamental Research Funds for the Central Universities, China (Grant No. 2024CDJXY022), Chinesisch-Deutsche Mobilit\"atsprogamm of Chinesisch-Deutsche Zentrum f\"ur Wissenschaftsf\"orderung (Grant No. M-0496).  Yisheng Chai acknowledged the support by the Natural Science Foundation of China (Grants NoS. 12227806, 12374081). Xiaolong Yang acknowledged the support by the Natural Science Foundation of China (Grant No. 12374038).

\end{document}